\documentclass{IEEEtran4PSCC}
\usepackage{amsmath}   
\usepackage{amsfonts}  
\usepackage{amssymb}   
\usepackage{subfigure}
\usepackage{tabularx}
\usepackage{xcolor, cite}
 \usepackage{url} 
\usepackage[normalem]{ulem}
\usepackage{float}
\usepackage{stfloats}
\ifCLASSINFOpdf
   \usepackage[pdftex]{graphicx}
\else
   \usepackage[dvips]{graphicx}
\fi

\makeatletter
\let\old@ps@headings\ps@headings
\let\old@ps@IEEEtitlepagestyle\ps@IEEEtitlepagestyle
\def\psccfooter#1{%
    \def\ps@headings{%
        \old@ps@headings%
        \def\@oddfoot{\strut\hfill#1\hfill\strut}%
        \def\@evenfoot{\strut\hfill#1\hfill\strut}%
    }%
    \def\ps@IEEEtitlepagestyle{%
        \old@ps@IEEEtitlepagestyle%
        \def\@oddfoot{\strut\hfill#1\hfill\strut}%
        \def\@evenfoot{\strut\hfill#1\hfill\strut}%
    }%
    \ps@headings%
}
\makeatother

\psccfooter{%
        \parbox{\textwidth}{\hrulefill \\ \small{24th Power Systems Computation Conference} \hfill \begin{minipage}{0.2\textwidth}\centering \vspace*{4pt} \includegraphics[scale=0.06]{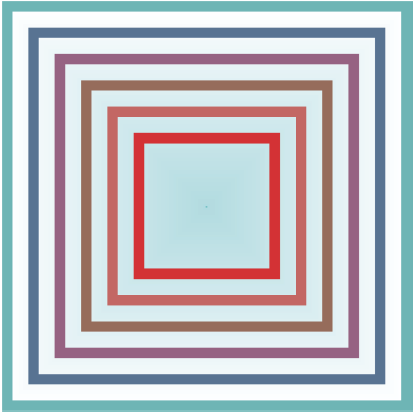}\\\small{PSCC 2026} \end{minipage} \hfill \small{Limassol, Cyprus --- June 8-12, 2026}}%
}

\begin{document}

\title{An OPF-based Control Framework for Hybrid AC-MTDC Power Systems under Uncertainty}

\author{
\IEEEauthorblockN{Hongjin Du, Rahul Rane, Weijie Xia, Pedro P. Vergara, Aleksandra Lekić}
\IEEEauthorblockA{Faculty of Electrical Engineering, Mathematics and Computer Science \\
Delft University of Technology\\
Delft, The Netherlands}
}


\maketitle

\begin{abstract}
The increasing integration of renewable energy, particularly offshore wind, introduces significant uncertainty into hybrid AC–HVDC systems due to forecast errors and power fluctuations. Conventional control strategies typically rely on fixed setpoints and neglect frequency deviations, which can compromise system stability under rapid renewable variations. To address this challenge, this paper presents a forecast-integrated, optimal power flow (OPF)-based adaptive control framework. Wind speed forecasts generated using a Random Forest model are incorporated into a time-coupled OPF to determine baseline converter setpoints in anticipation of wind fluctuations, which are further adjusted in real time based on actual operating conditions. An adaptive droop control scheme is developed that jointly considers DC voltage and AC frequency deviations. The effectiveness of the proposed control framework is validated through hardware-in-the-loop (HIL) simulations, demonstrating its capability to ensure stable and robust operation of hybrid AC–HVDC systems under high penetration of renewable energy.
\end{abstract}


\thanksto{\noindent This work is supported by the CRESYM project Harmony.}

\section{Introduction}

In recent years, the global energy landscape has been undergoing a steady transformation driven by the increasing adoption of renewable energy sources (RESs). Recent analyses, such as Ember’s Global Electricity Review 2025~\cite{ember2025review}, indicate that renewable energy has been playing an increasingly prominent role in global electricity generation over the past decade. Nevertheless, integrating high shares of RESs into large-scale power systems remains a significant technical and operational challenge \cite{REN21_2024}. 

High voltage direct current (HVDC) technology has been widely adopted as an efficient means of long-distance power transmission and renewable energy integration \cite{5681201}. Among different technologies, voltage source converter (VSC)-based HVDC, especially with modular multilevel converters (MMCs), has gained prominence due to its advanced controllability. Unlike traditional line-commutated converters (LCCs), VSCs can independently control active and reactive power, operate in weak or even passive grids, and support multi-terminal or meshed configurations. These features make VSCs particularly suitable for hybrid AC-HVDC power systems dominated by renewable energy. With the maturation of VSC–HVDC technology, the key challenge lies not in its technical feasibility but in ensuring secure and coordinated operation under various operating conditions, such as wind forecast uncertainties and grid disturbances.

In practice, these challenges are exacerbated by the intermittency and unpredictability of wind generation, which introduces power mismatches and additional operating costs in day-ahead and balancing markets \cite{rabiee2013stochastic}. The interaction between uncertain wind generation and the dynamics of DC systems further increases operational complexity. Existing studies typically address these mismatches by determining optimal operating points using stochastic or probabilistic optimal power flow (OPF)-based approaches \cite{atwa2011probabilistic, ROALD2023108725}. Although such methods better capture uncertainty than deterministic models, they typically impose high computational burdens and exhibit limited scalability, restricting their applicability in large-scale, real-time system operation.

Determining optimal operating points through OPF can guide the system toward economically efficient and coordinated operation. However, fixed set points derived from OPF are not suitable for dynamic operating conditions, as they cannot adapt to disturbances or forecast errors and may even lead to oscillations. Hence, maintaining secure operation requires appropriate control strategies.

Conventional voltage control strategies include master–slave control, which achieves accurate power sharing but depends on fast and reliable communication, and voltage droop control, which relies solely on local voltage and current measurements. By eliminating the need for extensive communication, droop control enables decentralized operation and improves scalability, making it particularly suitable for multi-terminal HVDC (MTDC) systems \cite{Beerten2011VSCMS, stojkovic2020adaptive}. However, fixed droop gains may result in uneven power sharing, which can lead to converter overloads or under-utilization \cite{wang2020dc}. To improve droop performance, adaptive and nonlinear droop control methods have been proposed in \cite{7967860, 9594778, Cao2013MinimizationOT, Zhang2021}. Nevertheless, existing control strategies for VSC–MTDC systems are developed under the assumption of constant frequency. As a result, droop strategies are primarily designed for DC voltage regulation, which neglects the inherent coupling between DC variables and AC frequency, as well as the influence of power injections at the point of common coupling (PCC) on overall system dynamics.

To address these limitations, this paper proposes a forecast-integrated, OPF-based adaptive control framework for AC–MTDC systems under high renewable penetration. The main contributions of this work are as follows: (1) Random Forest (RF) model-based wind forecasts are incorporated into a time-coupled OPF, capturing the temporal dynamics of renewable generation and enabling more informed operational planning; (2) an adaptive droop control strategy is developed, which dynamically adjusts droop coefficients in real time to enhance voltage and frequency stability and improve power-sharing accuracy across the network; and (3) the effectiveness of the proposed control framework is demonstrated through hardware-in-the-loop (HIL) simulations, highlighting its capability to maintain stable and robust operation of hybrid AC–MTDC systems under high renewable penetration.

The remainder of this paper is organized as follows. Section \uppercase\expandafter{\romannumeral2} presents the system modeling and problem formulation. Section \uppercase\expandafter{\romannumeral3} introduces the proposed forecast-integrated OPF and adaptive droop control framework. The feasibility of the proposed control scheme and its transient performance are evaluated through simulations in Section \uppercase\expandafter{\romannumeral4}, and Section \uppercase\expandafter{\romannumeral5} concludes the paper and outlines directions for future research.

\section{AC–MTDC System Modeling and Optimal Power Flow}

The hybrid AC–MTDC system considered in this study consists of multiple AC areas interconnected through a VSC-MTDC grid. The overall system model comprises detailed representations of the AC and DC networks, converters, and the OPF formulation used for coordinated operation analysis \cite{5589968}.

\subsection{AC System Modelling}

The AC subgrid is represented by steady-state nodal power flow equations that describe the active and reactive power injections at each bus. For each bus $i \in \mathcal{N}_{ac}$, the injected powers can be expressed as:
\begin{align}
P_{i} &= U_i \sum_{j=1}^{n} U_j \left[ G_{ij} \cos(\delta_i - \delta_j) + B_{ij} \sin(\delta_i - \delta_j) \right], \label{eq:ac_p}\\
Q_{i} &= U_i \sum_{j=1}^{n} U_j \left[ G_{ij} \sin(\delta_i - \delta_j) - B_{ij} \cos(\delta_i - \delta_j) \right], \label{eq:ac_q}
\end{align}
where $U_i$ and $U_j$ are the voltage magnitudes at buses $i$ and $j$, $\delta_i$ and $\delta_j$ are their corresponding voltage phase angles, and $G_{ij}$ and $B_{ij}$ denote the conductance and susceptance elements of the network admittance matrix, respectively.

At each bus, the active and reactive power balances are maintained as:
\begin{equation}
P_{Gi} - P_{Di} - P_{i} = 0, \quad
Q_{Gi} - Q_{Di} - Q_{i} = 0,
\label{eq:ac_balance}
\end{equation}
where $P_{Gi}$ and $Q_{Gi}$ represent the active and reactive power generation, and $P_{Di}$ and $Q_{Di}$ denote the corresponding power demands. 

Conventional generators are constrained by their operational limits:
\begin{equation}
P_{Gi}^{\min} \le P_{Gi} \le P_{Gi}^{\max}, \quad
Q_{Gi}^{\min} \le Q_{Gi} \le Q_{Gi}^{\max}.
\label{eq:ac_limits}
\end{equation}

In addition, the voltage magnitude and phase angle at each bus are constrained by their operational limits:
\begin{equation}
U_i^{\min} \le U_i \le U_i^{\max}, \quad
\delta_i^{\min} \le \delta_i \le \delta_i^{\max}, \quad \forall i \in \mathcal{N}_{ac}.
\label{eq:ac_voltage_limits}
\end{equation}

In hybrid AC–MTDC systems, converter stations are connected at specific AC buses, where their active and reactive power injections appear as additional terms in \eqref{eq:ac_balance}. These coupling points form the interface between the AC and DC subgrids, enabling coordinated control and optimization across the integrated system.

\subsection{DC System Modelling}

The DC subgrid is represented by nodal current balance equations that describe the relationship between DC voltages and line currents. For each DC bus $m \in \mathcal{N}_{dc}$, the injected current can be expressed as:
\begin{equation}
I_{dc,m} = \sum_{n=1, n \neq m}^{N_{dc}} Y_{dc,mn} (U_{dc,m} - U_{dc,n}),
\label{eq:dc_current_balance}
\end{equation}
where $U_{dc,m}$ and $U_{dc,n}$ denote the DC voltage magnitudes at buses $m$ and $n$, respectively, and $Y_{dc,mn}$ is the DC network admittance between them.

The injected DC power at bus $m$ is given by:
\begin{equation}
P_{dc,m} = U_{dc,m} I_{dc,m},
\label{eq:dc_power}
\end{equation}
where $P_{dc,m}$ is positive for power injection into the DC grid and negative for power withdrawal. $P_{dc,m}$ represents the DC-side power of the VSC connected at bus $m$. 

To ensure safe operation of the converters, the DC power at each terminal is constrained by its rated limits:
\begin{equation}
P_{dc,m}^{\min} \leq P_{dc,m} \leq P_{dc,m}^{\max}, \quad \forall m \in \mathcal{N}_{dc}.
\label{eq:dc_power_limits}
\end{equation}

\subsection{VSC Converter Modelling}

Converter station power losses, denoted as $P_{\text{loss}, k}$, are represented by three components: no-load losses, linear losses, and quadratic losses, which depend on the converter reactor current $I_{c, k}$. The converter losses and current relationships are modeled as:
\begin{align}
P_{\text{loss},k} &= a + b I_{c,k} + c I_{c,k}^2, \label{eq:Ploss}\\
P_{c,k}^2 + Q_{c,k}^2 &= 3U_{c,k}^2 I_{c,k}^2, \label{eq:current_relation}
\end{align}
where $P_{c,k}$ and $Q_{c,k}$ denote the active and reactive power at the converter AC terminal, and $U_{c,k}$ is the AC-side voltage magnitude of converter $k$. The loss coefficients are determined by conducting RTDS simulations over a range of power levels. The power losses in the converter are recorded for each case, and a second-order polynomial is fitted to the resulting data to derive the coefficients mentioned in Table~\ref{powerloss}.

\begin{table}[!htbp] 
    \centering
    \caption{VSC loss coefficients.}
    \footnotesize 
    \renewcommand{\arraystretch}{1.25}
    \begin{tabularx}{0.45\textwidth}{ >{\raggedright\arraybackslash}X >{\centering\arraybackslash}X >{\centering\arraybackslash}X >{\centering\arraybackslash}X }
     & \textbf{a [MW]} & \textbf{b [kV]} & \textbf{c [$\boldsymbol{\Omega}$]}\\ 
    \hline
        VSC 1 and 4 & 0.3183 & 1.4256 & 0.2049 \\
        VSC 2 and 3 & 0.1769 & 0.8594 & 0.3155 \\
    \hline
    \end{tabularx}
    \label{powerloss}
\end{table}

\subsection{Optimal Power Flow Formulation}
The system operating point is determined by solving an OPF problem to minimize generation cost while satisfying all AC, DC, and converter constraints. The OPF is formulated as:

\begin{equation}
\begin{aligned}
& \hspace{1em} \min_{x} \quad f(x) = \sum_{i \in \mathcal{G}} C_i(P_{Gi}) \\
\text{s.t.} &\quad \text{AC grid constraints (Section II-B),} \\
& \quad \text{DC grid constraints (Section II-C),} \\
& \quad \text{VSC converter constraints (Section II-D)}
\end{aligned}
\label{eq:opf_base}
\end{equation}

The generation cost of each generator is represented by a quadratic function:
\begin{equation}
C_i(P_{Gi}) = \alpha_i P_{Gi}^2 + \beta_i P_{Gi} + \gamma_i,
\end{equation}
where $\alpha_i$, $\beta_i$, and $\gamma_i$ are the cost coefficients corresponding to generator $i \in \mathcal{G}$.  

By solving this OPF problem, the optimal set of generation, DC voltages, and converter settings can be obtained, ensuring both economical operation and compliance with system constraints. This formulation provides the foundation for the forecast-integrated and adaptive droop control framework described in Section III.

\section{Forecast-Integrated OPF and Adaptive Droop Control Framework}
\subsection{Random Forest Model for Wind Speed Forecasting}

The forecasting task is formulated as a supervised multi-output regression problem.  
Let the training dataset be defined as:
\begin{equation}
\mathcal{D} = \big\{ (X_i, Y_i) \;\big|\; i = 1,2,\ldots,N \big\},
\end{equation}
where $X_i \in \mathbb{R}^d$ is the input feature vector and $Y_i \in \mathbb{R}^P$ is the output vector of $P$ future wind speed values.

A RF regressor is used and constructs an ensemble of $T$ decision trees $\{f_t(\cdot)\}_{t=1}^T$, each trained on a bootstrap sample of $\mathcal{D}$ \cite{lahouar2017hour, iannace2019wind}. For each tree, a subset of features is randomly selected at each split, ensuring diversity across the ensemble.

The prediction of a single tree is given by
\begin{equation}
\hat{Y}_i^{(t)} = f_t(X_i), \quad f_t: \mathbb{R}^d \to \mathbb{R}^P,
\end{equation}
and the RF prediction is obtained by averaging over all trees
\begin{equation}
\hat{Y}_i = \frac{1}{T} \sum_{t=1}^T \hat{Y}_i^{(t)}.
\end{equation}
Thus, the RF estimator can be written as
\begin{equation}
\hat{f}(X) = \frac{1}{T} \sum_{t=1}^T f_t(X),
\end{equation}
where $\hat{f}: \mathbb{R}^d \to \mathbb{R}^P$ denotes the ensemble predictor. In this study, the number of trees is set to $T=15$.

The forecasting model input combines historical wind-related variables and temporal features. Specifically, the last 48 time steps of wind power output and wind speed, together with the hour of the day encoded via sine and cosine transformations, are concatenated into a feature vector. The model outputs forecasts for the next 24 time steps, formulating a supervised multi-output regression task that maps the historical window to the future wind trajectory. Formally, given a history window of length $H=48$, the input vector is defined as:
\begin{equation}
\begin{split}
X_t = [&\, y_{t-H}, \ldots, y_{t-1}, \; s_{t-H}, \ldots, s_{t-1}, \\ 
      &\, \sin\!\Big(\frac{2\pi h_t}{24}\Big), \; \cos\!\Big(\frac{2\pi h_t}{24}\Big) \, \Big]
\end{split}
\end{equation}
where $y$ denotes wind output, $s$ denotes wind speed, and $h_t$ is the hour of day at time $t$.  
The corresponding output is the sequence of future wind speeds
\begin{equation}
Y_t = \big[ s_{t}, s_{t+1}, \ldots, s_{t+P-1} \big],
\end{equation}
where $P=24$ represents the prediction horizon.

\begin{figure}
    \centering
    \includegraphics[width=0.8\linewidth]{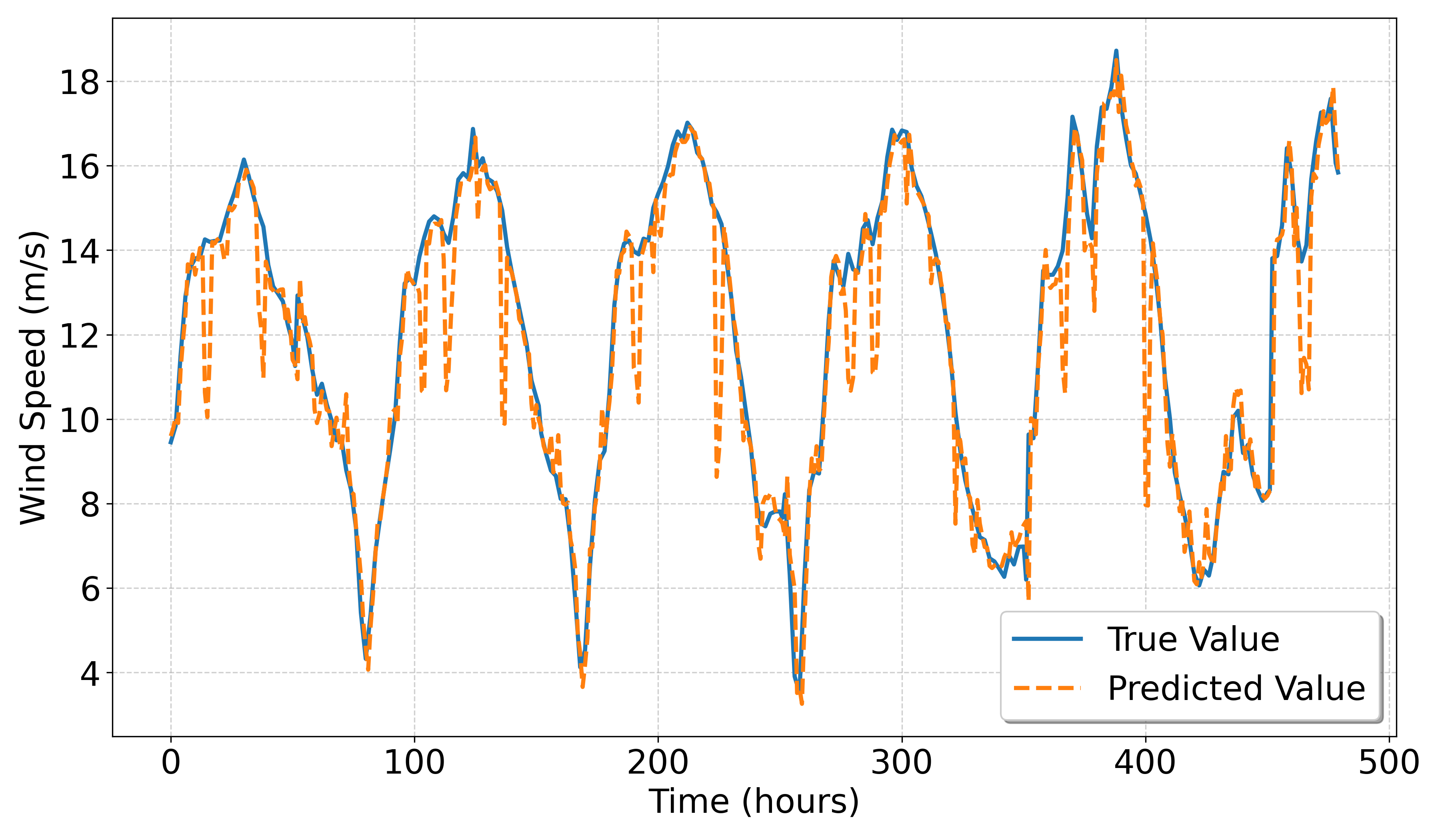}
    \caption{Wind speed prediction results.}
    \label{windspeedpre}
\end{figure}

The RF model is trained on these input–output pairs. During inference, the trained model generates predictions $\hat{Y}_t$ for unseen test samples, thereby providing a 24-step ahead wind speed forecast. Figure~\ref{windspeedpre} illustrates the comparison between the observed and predicted wind speeds using the RF model. The results demonstrate that the model effectively reproduces the temporal dynamics of wind speed with a high degree of accuracy. The dataset used in this study is provided by the IEEE DataPort “Hybrid Energy Forecasting and Trading Competition” \cite{heft2024}. 
\subsection{Adaptive Droop Control}

For fixed droop control, the droop coefficients are usually dependent on the respective converter ratings. However, powers and DC voltage deviations of converters with fixed droop control may reach their limits in case of unexpected events such as faults or the disconnection of a station. This problem can be mitigated using an adaptive droop scheme. 

Various adaptive voltage droop control methods have been proposed over the years. The power headroom (i.e., the difference between the rated capacity and the present loading) was introduced to improve power sharing \cite{6242419}, but the DC voltage deviation during both transient and steady-state conditions was not considered. In \cite{8373718}, an adaptive droop control strategy addressing both power sharing and DC voltage deviation was introduced. However, the method neglects frequency dynamics, which are essential for coordinated AC–DC operation.

In this study, an adaptive droop control scheme is proposed that separately considers DC voltage and AC frequency deviations. 
The active power reference of converter~$k$ is defined as:
\begin{equation}
    P_{c,k} = P_{c,ref} + \Delta P_{v,k} - \Delta P_{f,k},
    \label{f_droop}
\end{equation}
where $\Delta P_{v,k}$ and $\Delta P_{f,k}$ represent the active power adjustments associated with DC voltage and AC frequency regulation at converter~$k$, respectively.

The voltage-related adjustment is expressed as:
\begin{equation}
    \Delta P_{v,k} = \frac{U_{dc,\text{ref}} - U_{dc,k}}{k_{v,k}},
\end{equation}
and the frequency-related adjustment is defined by:
\begin{equation}
    \Delta P_{f,k} = \frac{f_{k,\text{ref}} - f_k}{k_{f,k}},
\end{equation}
where $k_{v,k}$ and $k_{f,k}$ are the voltage and frequency droop coefficients, respectively. These coefficients determine the converter’s sensitivity to local DC voltage and frequency deviations.

The droop coefficients are determined according to the converter’s available power margin and the corresponding operating range. For voltage control, the coefficient is given by:
\begin{equation}
    k_{v,k} = \frac{U_{dc,k}^{\max} - |U_{dc,ref} - 1|}{\alpha_v (P_k^{\max} - |P_k|)}.
\end{equation}
Similarly, the frequency droop coefficient is formulated as:
\begin{equation}
    k_{f,k} = \frac{\Delta f_{k}^{max}}{\alpha_f (P_k^{\max} - |P_k|)},
\end{equation}
where the $\Delta f_{k}^{max}$ is the maximum allowable permissible frequency deviation. $\alpha_v$ and $\alpha_f$ are tunable sensitivity parameters introduced to dynamically adjust the converter’s contribution to voltage and frequency regulation. For example, this mechanism can allocate a greater proportion of available active power to frequency recovery while temporarily relaxing voltage regulation during severe disturbances.

\subsection{Control Framework}

\begin{figure}[htbp]
\centering
\centerline{\includegraphics[width=0.43\textwidth]{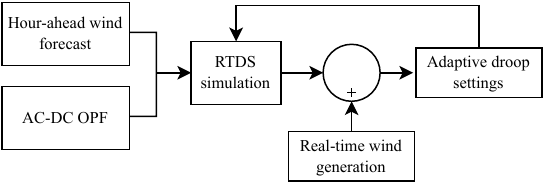}}
\caption{Flowchart of the proposed control strategy.}
\label{flowchart}
\end{figure}

Building on the OPF formulation presented above, the proposed control framework integrates wind power forecasting with adaptive droop control for hybrid AC--MTDC systems incorporating offshore wind farms. Wind speed forecasts generated using a Random Forest model are embedded into a time-coupled OPF to determine the baseline VSC droop settings in anticipation of wind fluctuations. Subsequently, the VSC parameters are dynamically tuned in real time according to the actual wind conditions. The overall process, highlighting the interaction between forecasting, OPF computation, and adaptive droop implementation, is illustrated in Fig.~\ref{flowchart}.

\begin{figure}[htbp]
    \centering
    \centerline{\includegraphics[width=0.38\textwidth]{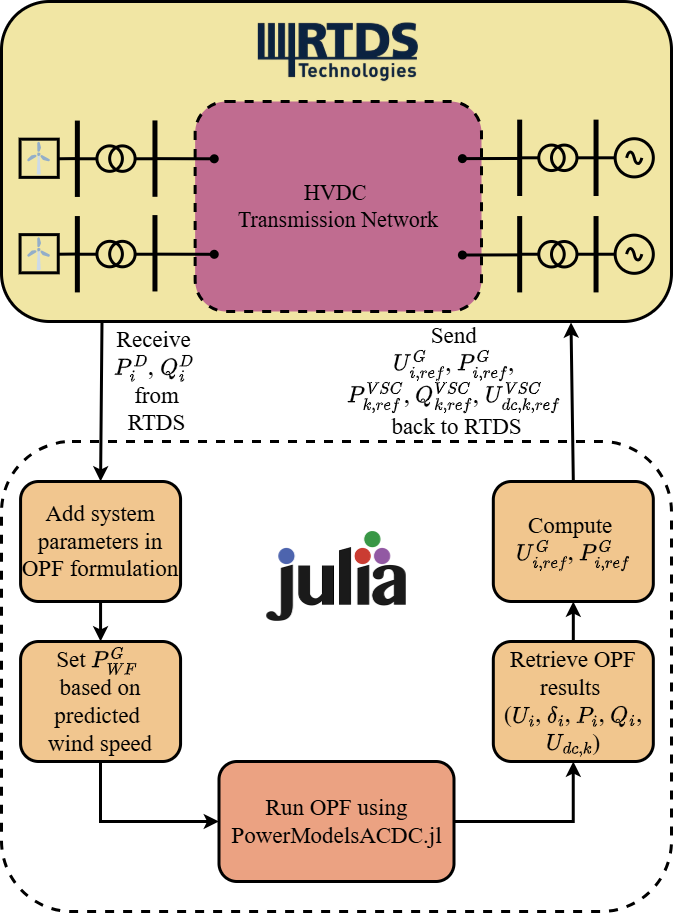}}
    \caption{Computational flow within the HIL control framework.}
    \label{opf_flow}
\end{figure}

The computational flow within the proposed HIL control framework is illustrated in Fig.~\ref{opf_flow}. The Real-Time Digital Simulator (RTDS) exchanges real-time data with an external optimization framework based on PowerModelsACDC.jl developed in Julia \cite{8636236}, where the Interior Point OPTimizer (IPOPT) is employed to solve the OPF problem. 

Specifically, RTDS transmits system measurements such as the active and reactive power demands $(P_i^D, Q_i^D)$ to the Julia environment. The predicted wind speed is used to determine the wind farm generation $(P_{WF}^G)$, which is included in the OPF formulation together with system parameters and operational constraints $(U_{i,\min}, U_{i,\max}, P_{i,\max}^G, \ldots)$. After executing the OPF, the computed results, including bus voltages, phase angles, power injections, and DC link voltages $(U_i, \delta_i, P_i, Q_i, U_{dc,i})$, are retrieved to calculate updated setpoints $(U_{i,\mathrm{ref}}^G, P_{i,\mathrm{ref}}^G, P^{VSC}_{k,\mathrm{ref}}, Q^{VSC}_{k,\mathrm{ref}}, U^{VSC}_{k,\mathrm{ref}}, U^{VSC}_{dc,k,\mathrm{ref}})$. These setpoints are then communicated back to RTDS for real-time implementation. The data exchange between RTDS and Julia is achieved through the GTNETx2 communication interface card, while system modeling and monitoring are carried out using RSCAD.

\section{Simulation Results}

\begin{figure*}[!htbp]
    \centering
    \subfigure[IEEE 9 bus system]{
        \includegraphics[width=1\columnwidth]{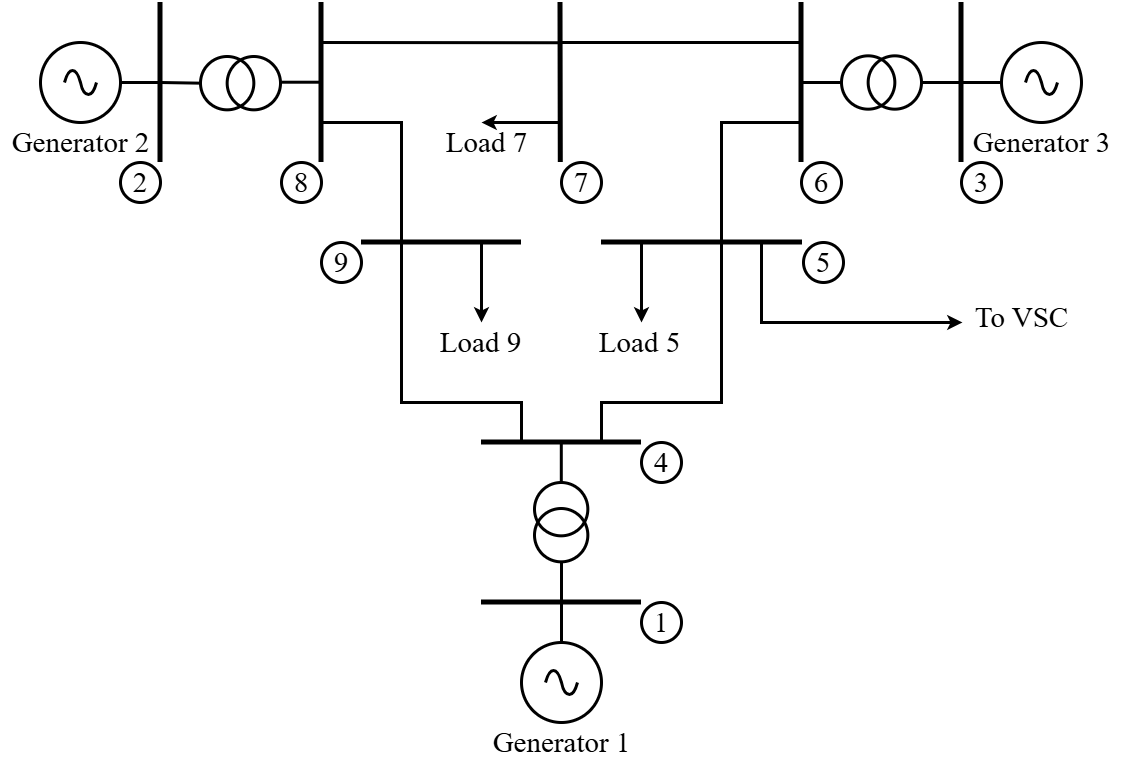}}%
    \subfigure[4 terminal HVDC test system]{
        \includegraphics[width=\columnwidth]{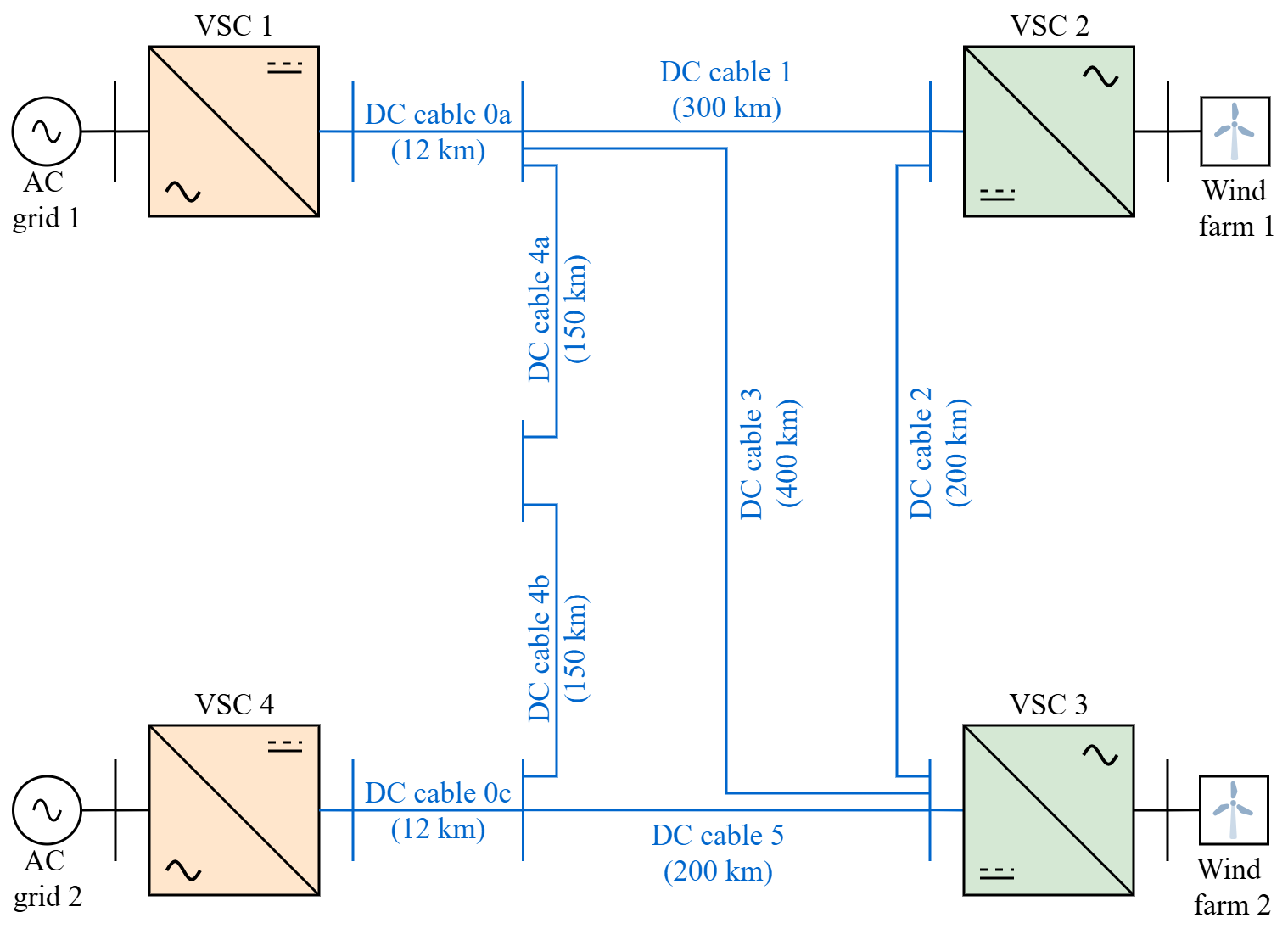}}
    \caption{Configuration of a 4-terminal AC-MTDC system where the AC grid 1 and 2 represent the IEEE 9 bus system.}
    \label{topology}
    \vspace*{-5mm}
\end{figure*}

To evaluate the effectiveness of the proposed control framework, a four-terminal VSC-MTDC system is adopted as the test platform. As illustrated in Figure~\ref{topology}, the system interconnects two IEEE 9-bus benchmark AC grids and two offshore wind farms through a DC network. The system parameters are based on the data provided in \cite{shetgaonkar2024enhanced} and are summarized in Table~\ref{parameters}. 

\begin{table}[!htbp]
    \centering
    \caption{Main parameters of the hybrid AC-MTDC test system.} \label{parameters}
    \renewcommand{\arraystretch}{1.5}
    \begin{tabular}{p{3cm} c c}
        \hline
        System parameter & Onshore MMC & Offshore MMC\\ 
        \hline
        Rated converter power & 2000 MVA & 2000 MVA \\
        Fundamental frequency & 60 Hz & 50 Hz \\
        AC grid voltage  & 230 kV & 66 kV \\
        DC grid voltage (bipolar) & $\pm$525 kV & $\pm$525 kV \\
        Transformer reactance & 0.18 p.u. & 0.15 p.u. \\
        MMC arm inductance & 0.05 mH & 0.1 mH \\
        MMC arm resistance & 0.157~$\Omega$ & 0.157~$\Omega$ \\
        Submodule capacitance & 25 mF & 15 mF \\
        Submodules per arm & 220 & 220\\
        \hline
    \end{tabular}
\end{table}

In this paper, the proposed control strategy is implemented and compared against conventional active power control, DC voltage control, and adaptive voltage droop control to evaluate system performance under various operating conditions. For all control strategies, VSCs 2 and 3 operate in AC voltage and frequency (V–f) control mode, while the control modes applied to VSCs 1 and 4 are defined as follows:
\begin{enumerate}
    \item \textbf{Active power control:} VSC 1 operates under active power control and VSC 4 under DC voltage control, without adaptive response to voltage or frequency fluctuations. 
    \item \textbf{DC voltage control:} VSC 1 operates under DC voltage control mode and VSC 4 under active power control, without adaptive response to voltage or frequency fluctuations. 
    \item \textbf{Adaptive voltage droop control:} VSC 1 and VSC 4 operate under voltage droop control. Unlike conventional droop schemes with fixed coefficients, the droop gains ($k_v$) are dynamically updated according to system operating conditions.
    \item \textbf{Proposed droop control:} VSC 1 and VSC 4 operate under proposed droop control. Similar to adaptive droop control, the droop coefficients ($k_v$ and $k_f$) are dynamically updated according to system operating conditions.
\end{enumerate}

Three representative scenarios are designed to evaluate the control strategies, reflecting realistic operating conditions and progressively increasing disturbance severity to assess control robustness.
\begin{enumerate}
    \item \textbf{Scenario 1 — Normal condition:}  
    The system operates under nominal conditions while subjected to realistic wind forecast uncertainty in the two offshore wind farms. 
    \item \textbf{Scenario 2 — AC Step:}  
    At $t = 1\,$s, a sudden load increase of 500 MW is applied at bus 5 in the AC network connected to VSC 1.
    \item \textbf{Scenario 3 — DC Disturbance:}  
    At $t = 1\,$s, a pole-to-pole DC fault occurs at the center of DC cable 1, followed by permanent disconnection of the affected line after fault clearance. 
\end{enumerate}

It should be noted that since the active power control mode and DC voltage control mode correspond to complementary actions (and vice versa), their responses are symmetric. Therefore, only the performance of VSC 1 is presented for all scenarios to avoid redundancy.

\begin{figure}[t]
    \centering
    \includegraphics[width=0.43\textwidth]{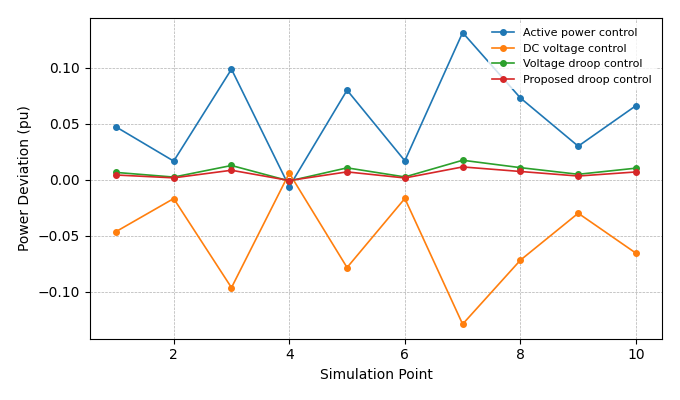}
    \caption{Power deviation at VSC 1.}
    \label{p_normal_1}
\end{figure}
\begin{figure}[!htbp]
    \centering
    \subfigure[Active power performance]{
        \includegraphics[width=0.43\textwidth]{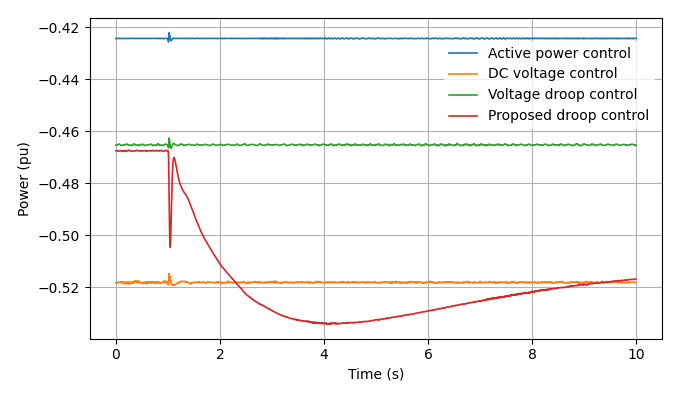}}
    \subfigure[DC voltage performance]{
        \includegraphics[width=0.43\textwidth]{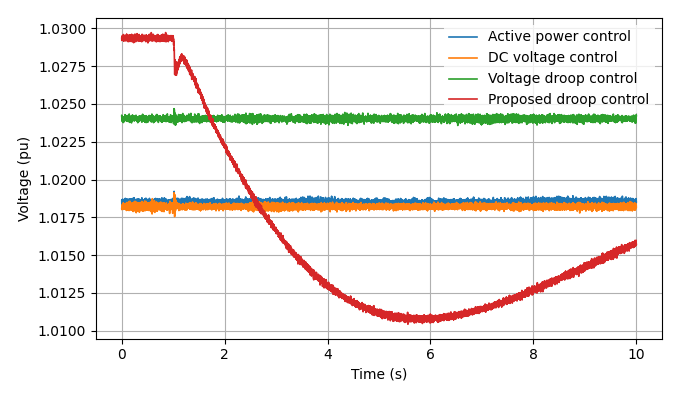}}
    \subfigure[Frequency performance]{
        \includegraphics[width=0.43\textwidth]{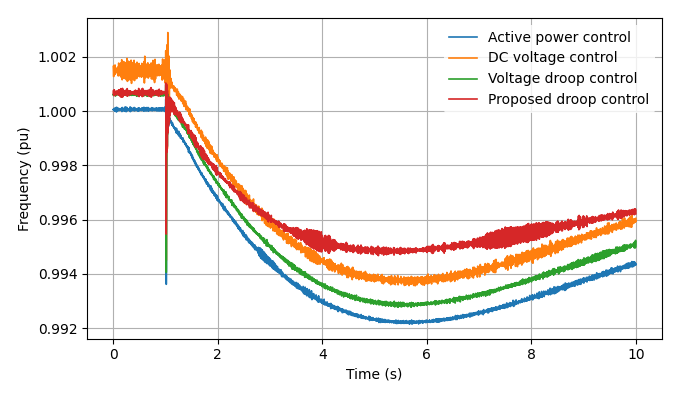}}
    \caption{Dynamic responses of VSC 1 under different control strategies in Scenario 2.}
    \label{ac_step}
    \vspace*{-5mm}
\end{figure}

In Scenario 1, the system operates under steady-state conditions using set points obtained from OPF calculations based on predicted wind speeds, while the actual wind speeds are applied in the simulation. The responses of the three control strategies are evaluated, and the active power deviations from the OPF results under actual wind speeds at VSC~1 are illustrated in Figure~\ref{p_normal_1}. The results indicate that the proposed droop control yields the smallest active power deviations, followed by adaptive voltage droop control, while active power control and DC voltage control exhibit the largest deviations. This demonstrates that incorporating adaptive mechanisms that respond to both voltage and frequency deviations effectively mitigates the impact of wind forecast errors on converter power dispatch.  

\begin{figure}[t!]
    \centering
    \subfigure[Active power performance]{
        \includegraphics[width=0.43\textwidth]{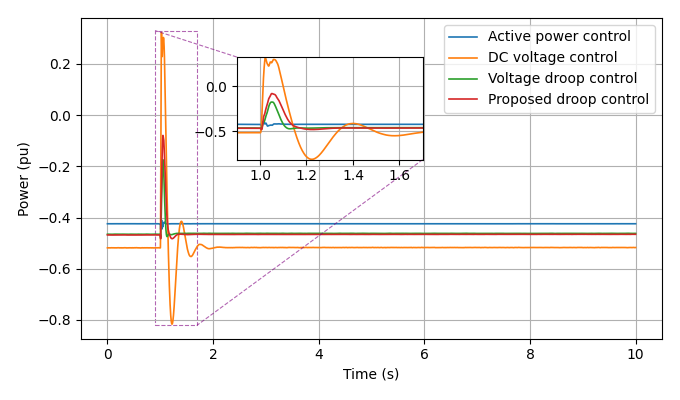}}
    \subfigure[DC voltage performance]{
        \includegraphics[width=0.43\textwidth]{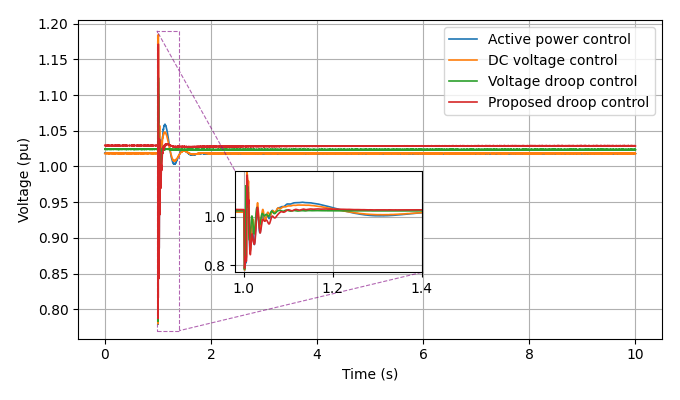}}
    \subfigure[Frequency performance]{
        \includegraphics[width=0.43\textwidth]{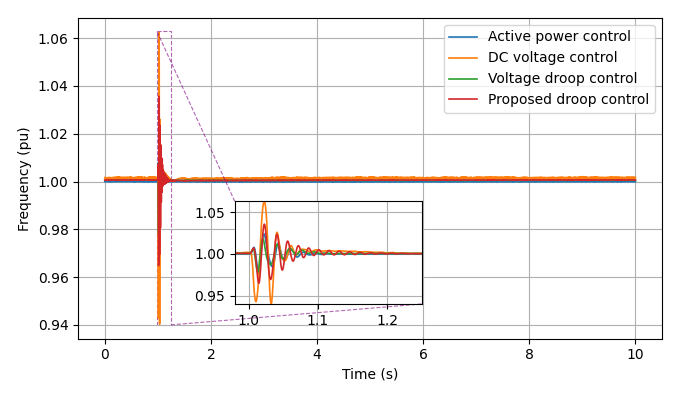}}
    \caption{Dynamic responses of VSC 1 under different control strategies in Scenario 3.}
    \label{dc_line}
\end{figure}

Figure~\ref{ac_step} illustrates the dynamic response of VSC 1 under a 500 MW load increase in AC grid 1 at $t = 1$ s, comparing four control strategies. Both active power control and DC voltage control exhibit the frequency deviations and slow recovery, as their fixed power references prevent adaptive response to system disturbances. The adaptive voltage droop control introduces voltage-dependent adjustment, which slightly improves DC voltage stability but remains inadequate for mitigating frequency excursions on the AC side. This limitation arises because its regulation is driven solely by DC voltage deviations, neglecting frequency dynamics during AC-side disturbances. In contrast, the proposed adaptive droop control, as defined in \eqref{f_droop}, dynamically coordinates voltage and frequency deviations within the active power adjustment process. This enables a balanced response between DC and AC dynamics, effectively maintaining both DC voltage and system frequency close to their nominal values. Consequently, the frequency dip is notably less severe, and the recovery is faster and smoother, demonstrating improved transient stability and power-sharing capability compared to other methods.

Figure~\ref{dc_line} presents the dynamic responses of VSC 1 in Scenario 3, where a DC line fault occurs. In this case, both active power control and DC voltage control again exhibit poor disturbance tolerance, showing large voltage deviations and delayed power recovery due to their static set points. The adaptive voltage droop control outperforms these two by actively adjusting power output in proportion to voltage deviations, thereby enhancing post-fault voltage restoration. The proposed droop control demonstrates performance very close to that of the voltage droop control while clearly surpassing active power control and DC voltage control in terms of overall system stability. Although the voltage droop control achieves faster local voltage restoration inherently, the proposed method provides a more balanced dynamic response by coordinating both voltage and frequency regulation.

Overall, the proposed scheme yields the most balanced and robust transient performance, enhancing stability and system resilience across diverse disturbance conditions.

\section{Conclusion}

This paper proposes a forecast-integrated OPF-based control framework incorporating a proposed adaptive droop control strategy for hybrid AC–MTDC systems with offshore wind farms. Wind power forecasts generated using a Random Forest model are embedded directly into a time-coupled OPF, enabling dynamic adjustment of VSC droop settings in response to anticipated wind fluctuations. Case studies on a four-terminal AC-MTDC test system demonstrate that the proposed control method effectively mitigates active power deviations under normal operating conditions with wind forecast uncertainty, maintaining power dispatch close to optimal set points. During sudden AC-side disturbances, such as large load steps, it achieves faster and smoother recovery of both DC voltage and system frequency compared to active power control, DC voltage control, and adaptive voltage droop control. In the event of DC line faults, the proposed method exhibits performance closely matching adaptive voltage droop control while clearly outperforming active power and DC voltage control, providing a more balanced dynamic response across the system. Future work will focus on extending the framework to larger multi-region AC–DC systems and exploring reinforcement learning-based strategies for adaptive VSC control to further improve system performance and coordination.

\bibliographystyle{IEEEtran}
\bibliography{manual.bib}   

\end{document}